\documentclass[a4paper]{jpconf}
\usepackage{amsmath}
\usepackage{wasysym}
\usepackage{amssymb}
\usepackage{bm}
\usepackage{graphicx}
\usepackage{feynmf}
\usepackage{siunitx}
\begin{document}

\title{Scientific Opportunities at the ARIEL Electron Linac}
\author{Jan Bernauer$^{1,2}$, Ross Corliss$^1$, Susan Gardner$^3$, Michael Hasinoff$^4$, \\ Rituparna Kanungo$^5$, Jeffery Martin$^6$, Richard Milner$^7$,\\  Katherine Pachal$^8$, Toshimi Suda$^9$ and Stanley Yen$^8$}
\date{June 2022}

\address{$^1$ Stony Brook University, Stony Brook, NY, USA\\
$^2$ RIKEN BNL Research Center, Upton, NY, USA\\
$^3$ University of Kentucky, Lexington, KY, USA \\ 
$^4$ University of British Columbia, Vancouver, BC, Canada \\ $^5$ St.\ Mary's University, Halifax, NS, Canada \\ $^6$ University of Winnipeg, Winnipeg, MB, Canada \\ $^7$ Massachusetts Institute of Technology, Cambridge, MA, USA \\ $^8$ TRIUMF, Vancouver, BC, Canada \\ $^9$ Tohoku University, Sendai, Japan}

\ead{ross.corliss@stonybrook.edu}

\begin{abstract}
This paper gives an overview of the scientific opportunities at the ARIEL electron accelerator identified in open discussion at the workshop, including applications in hadron structure, astrophysical processes, tests of quantum electrodynamics, dark matter and other BSM physics, and material science.  
\end{abstract}

\section{Introduction}
\bigskip

The Advanced Rare Isotope Laboratory (ARIEL), now under construction, will extend TRIUMF's existing isotope program, enabling world-class research on the nature of atomic nuclei, the origin of the heavy chemical elements, and contributing more broadly to nuclear medicine, materials science, and nuclear and particle physics research.  ARIEL is driven jointly by the TRIUMF cyclotron and by a newly constructed superconducting electron accelerator (e-linac) which will enable isotope production via photo-production and photo-fission.  Designed to provide a continuous, 10~mA beam with energies up to 30~MeV (300~kW), and with several upgrade paths available (see T.\ Planches's contribution to these proceedings for details), this e-linac can also enable new physics programs independently of the ISAC facility, both in its current form and through those upgrades.

In the open discussion sessions of the May, 2022 workshop, a number of possible new scientific opportunities at the e-linac were identified. We have collected them here with the purpose of providing a starting point for the interested reader to consider them in more depth. 

\section{Hadron Structure and Astrophysics}
\subsection{Low Q$^2$ Elastic Electron Scattering}
Non-perturbative Quantum Chromodynamics describes the physics inside nucleons and nuclei, with many of the bulk properties emerging from the dynamics of the system instead of being determined by the properties of the constituents---in contrast, for example, to the QED description of atoms.
Elastic electron-hadron scattering allows us to directly access these intrinsic, emergent properties, validating our current theoretical model. Precision measurements are benchmarks for lattice calculations and provide important input to other fields, from astrophysics to atomic physics.

\subsubsection{Proton Targets}
Using electron-proton scattering on a hydrogen-containing target, one can determine the elastic form factors $G_E$ and $G_M$, which describe the distribution of charge and currents inside the proton. Intrinsic properties of the proton can be related to these, such as the charge and magnetic radii of the proton, which are related to the slope of these form factors at zero four-momentum transfer. Other emergent phenomena, such as a meson cloud that forms around the bare nucleon, leave signatures in the shape of the form factors as well. 

To first order, the elastic electron-proton scattering cross section can be written as  
\begin{equation}
    \frac{\text{d}\sigma}{\text{d}\Omega}=\frac{1}{\epsilon(1+\tau)}\left[\epsilon G_E^2+\tau G_M^2\right]\left(\frac{\text{d}\sigma}{\text{d}\Omega}\right)_\textrm{Mott}\label{eq:redcross},
\end{equation}
where the photon polarization $\epsilon$ varies between 0 and 1 for backward to forward scattering, and $\tau=Q^2/(4m_\textrm{proton}^2)$ is proportional to the exchanged four-momentum squared.  The electric and magnetic radii are given by
\begin{equation}
    \left<r_{E(M)}\right>^2=-\frac{6\hbar^2}{G_{E(M)}(0)}\left.\frac{\text{d}G_{E(M)}}{\text{d}Q^2}\right|_{Q^2=0},
\end{equation}
in both cases requiring an extrapolation of the measured form factor to $Q^2=0$.

The charge radius also plays a role in spectroscopy as part of the Lamb-shift. It has attracted interest over the last decade, in form of the so-called proton radius puzzle, originally established in 2010 by a \SI{4}{\percent} difference between extractions using muonic spectroscopy \cite{pohlSizeProton2010} ($r_\text{p}=\SI{0.84184(67)}{\femto\meter}$) and both the results of the Mainz high-precision form factor experiment \cite{bernauerHighprecisionDeterminationElectric2010} ($r_\text{p}=0.879(5)_\text{stat}(6)_\text{syst}\,\si{\femto\meter}$)  and the CODATA value \cite{mohrCODATARecommendedValues2008} ($r_\text{p}=\SI{0.8768(69)}{\femto\meter}$), based on a series of normal hydrogen spectroscopy measurements and radius extractions from earlier scattering data.

\begin{figure}[htb]
    \centering
    \includegraphics[width=0.7\textwidth]{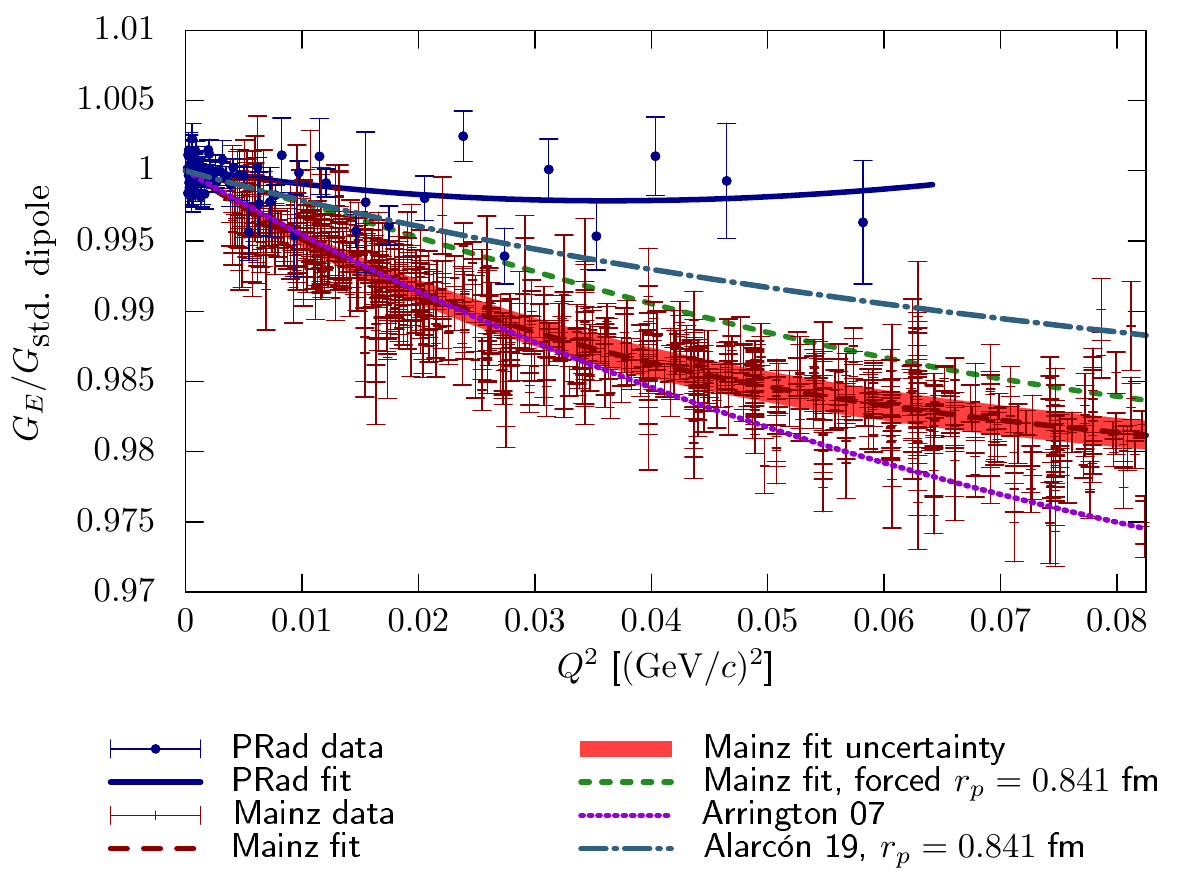}
    \caption{Recent results for $G_E$, normalized to the standard dipole. Red data points are the Mainz data \cite{bernauerHighprecisionDeterminationElectric2010} with fit and error band. Additionally, a fit to the Mainz data with a radius forced to the small value from muonic spectroscopy is shown (green, dashed). The PRad data and fit \cite{Xiong:2019umf} are shown in blue. The lilac dashed line labeled ``Arrington 07'' is a fit to pre-Mainz data \cite{arrington2007}. The teal dot-dashed line is from \cite{Alarcon:2018zbz}, a dispersively improved effective-field-theory calculation with the radius as the only free parameter, chosen here to be the value from muonic spectroscopy. The general agreement between the Mainz and earlier data, the PRad data, and this calculation is poor, raising questions about the reliability of all existing form factor measurements.}
    \label{fig:gewhatweknow}
\end{figure}

Recent measurements, especially in spectroscopy, tend to agree with the smaller radius, however some measurements prefer the larger. In the regime of scattering, the recent PRad \cite{Xiong:2019umf} result gives the smaller result but exhibits a clear difference from all earlier measurements not only in the extrapolation to $Q^2=0$, but also over the whole overlap with earlier measurements (see Fig.~\ref{fig:gewhatweknow}), putting our knowledge about the proton form factors, not only at small $Q^2$, into question.

\begin{figure}[htb]
    \centering
    \includegraphics[width=0.7\textwidth]{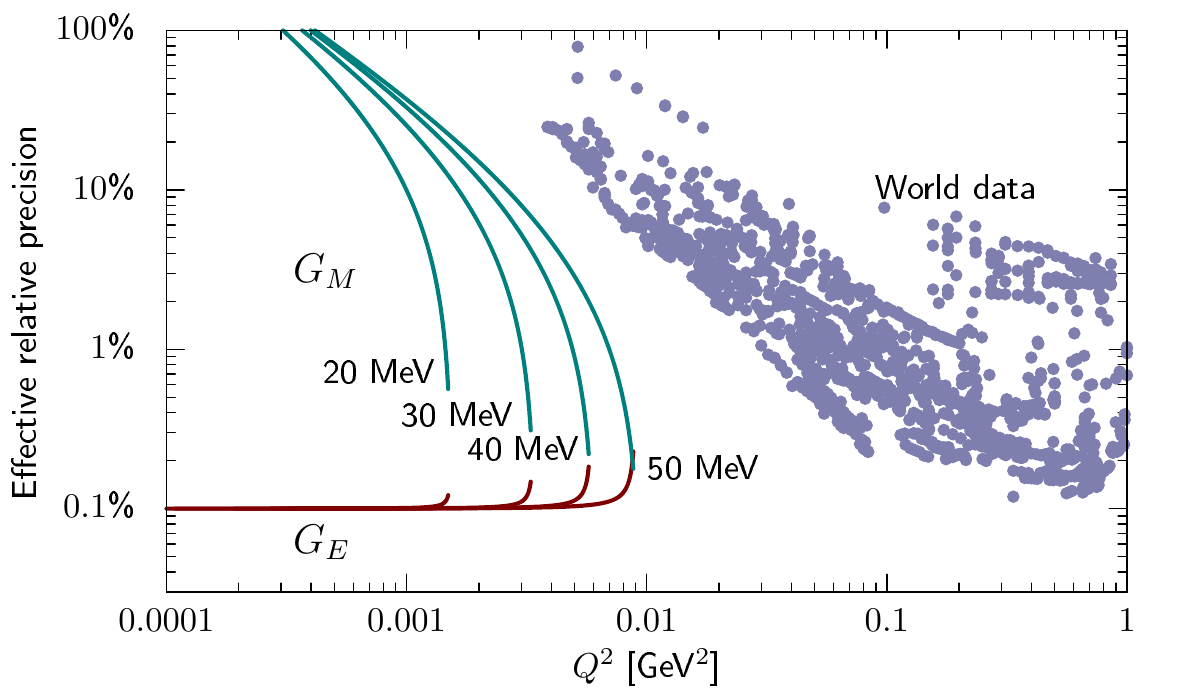}
    \caption{Effective relative precision of the form factors from cross section measurements. Teal ($G_M$) and red ($G_E$) lines show the reach and sensitivity for DarkLight@ARIEL assuming cross section measurements on the 0.2\% level. Blue points represent the precision on $G_M$ for existing data \cite{PhysRevC.90.015206}. Precise data on the form factors at $Q^2$ below 0.1 GeV$^2$ are required to extract a reliable radius with minimal model dependence.}
    \label{fig:wwkgm}
\end{figure}

To minimize the effect of the extrapolation, experiments try to reach smaller and smaller $Q^2$. PRad does so by instrumenting extremely small scattering angles at comparatively large beam energies. A complementary measurement would measure the form factors in the same $Q^2$ range by instrumenting larger scattering angles using smaller beam energies. This would also allow the measurement of backward angles and the extraction of the magnetic radius; since the contribution of $G_M$ is suppressed by $\tau\propto Q^2$, data is sensitive to $G_M$ only if $\varepsilon$ is small, i.e., if $\theta$ is close to 180$^\circ$. 
Currently, data at low $Q^2$ are measured at higher beam energies, moving data sensitive to $G_M$ to larger $Q^2$. Therefore any extraction relies on the stiffness of the fit to extrapolate from larger $Q^2$. The Mainz fit \cite{bernauerHighprecisionDeterminationElectric2010}, which is relatively flexible, finds some structure at $Q^2=0.03\,(\mathrm{GeV}/c)^2$, while other fits with less flexibility do not resolve this structure. This leads to large variations in the extracted value for the magnetic radius. Similar to the charge radius, the low-$Q^2$ shape of the magnetic form factor is important for spectroscopy via the so-called Zemach-radius. A reliable determination of the magnetic radius and low-$Q^2$ magnetic form factor shape would therefore be another connection to atomic physics. The sensitivity of current data to $G_M$ as well as the possible reach of an experiment at ARIEL are shown in Fig.~\ref{fig:wwkgm}.

The large beam current possible at ARIEL would permit a dramatically thinner target while maintaining sufficient luminosity, even at backward angles. As part of the MAGIX program, the A1 and MAGIX collaborations are developing a jet target \cite{Schlimme:2021gjx} which would realize such a thin target, potentially without any background from cell walls or similar. Additionally, the beam-jet interaction region is essentially point-like, reducing systematic uncertainties stemming from acceptance shifts for extended targets and from external radiation.

\subsubsection{Stable Nuclei}

In addition to a proton target, it is possible to extend such a program to light nuclei such as $^2$H, $^3$H, $^3$He, $^4$He and lithium---nuclei which are also the focus of muon spectroscopy, with some results already available (see \cite{Antognini:2021icf} and references therein) and others planned. This would open the way to additional cross checks between spectroscopic and accelerator-based measurements.

Additionally, measurements of low-$Q^2$ form factors for solid materials like carbon are almost trivial, but could improve the world data set considerably.

\subsection{Electron Scattering from Radioactive Nuclei}

The production of radioactive isotopes and the study of their structures and reactions are the primary objectives of TRIUMF's  facility. With its 300~kW superconducting electron driver, the ARIEL facility will expand the isotope production capabilities and further increase its research opportunities at TRIUMF.

In addition to isotope production, there is the potential to develop a program to perform electron scattering on these isotopes, which would take full advantage of ARIEL's powerful electron driver with TRIUMF's ISOL (Isotope Separation On-Line) facilities and to differentiate research opportunities in the field of nuclear physics from the other radioactive-isotope (RI) facilities, especially fragmentation facilities, in operation and under construction worldwide.

Although such electron scattering work has played an essential role in structure studies of atomic nuclei, it has never been applied to exotic nuclei to date. Studies of the structure and reaction of exotic isotopes have, without exception, been conducted with strongly-interacting particles.

Discussions on the possible application of electron scattering to radioactive nuclei started in the late 1990s~\cite{Katayama2003}, when powerful heavy-ion accelerator facilities began to be planned, assuming a collider scheme consisting of an electron ring and a large-scale storage ring to store an energetic radioactive-nucleus beam.

The world's first electron scattering facility dedicated to short-lived RIs, the SCRIT facility at RIKEN in Japan, is in operation today.  It is not a collider, but instead consists of an electron storage ring and an ISOL, based on a novel ion-trapping technique named SCRIT (Self-Confining Radioactive-isotope Ion Target)~\cite{Wakasugi2004}. The SCRIT technique achieves a high luminosity, $\sim 10^{27} \mathrm{cm}^{-2} \mathrm{s}^{-1}$ with a very small number of ions, typically $10^9$~\cite{Tsukada2017}, whose luminosity is comparable with those expected at the collider projects. This luminosity allows measuring the charge form factor of medium-heavy nuclei covering the first diffraction maximum, and one can determine, in addition to their charge radii, the size and shape of exotic nuclei~\cite{Suda2017}. For more information on the facility, see T. Suda's contribution to these proceedings. 

The radioactive isotopes at the SCRIT facility are produced at an ISOL for the photo-fission reaction of $^{238}$U, but the driver beam power will be limited to a maximum of a few kW due to regulations. This SCRIT scheme looks quite compatible with the ARIEL facility, and the ARIEL driver power is more than two orders of magnitude higher.  

If an electron storage ring with a few 100~MeV electron beam energy is equipped at ARIEL, the powerful SC electron driver of ARIEL will significantly expand accessible RIs for electron scattering compared with those at the SCRIT facility.

\subsection{Photonuclear Reactions for Radioactive Isotopes}

Photonuclear reactions provide a way to study the structure of nuclear excited states with well-understood transition operators.  Indeed, the photo-absorption cross sections so far available for many stable nuclei~\cite{Berman1975} provided crucial information to the current understanding of the structures of excited states including collective excitation dynamics. 

For radioactive nuclei, the Coulomb excitation in a nucleus-nucleus collision at a forward angle has been the only way so far to study their photonuclear responses. Unfortunately, its interpretation has been always questioned due to the fact that it is not a purely electromagnetic reaction. 

Installing an electron scattering facility as discussed above at ARIEL would also be the first to make possible the measurement of the total photoabsorption cross section of radioactive nuclei with a purely electromagnetic probe. As is well known, the inelastic electron-scattering cross section at the ultra-forward angle, $\theta \sim 0 ^\circ$, is related to the photo-nuclear cross section by virtual-photon theory~\cite{Carlos2007}, exactly in the same manner as the Coulomb excitation in a
nucleus-nucleus collision.

Detecting scattered electrons that have lost energy in inelastic electron scattering off exotic nuclei at a very forward angle allows for determining the total photo-absorption cross section.  One can cover fully the Giant Dipole Resonance (GDR) region with an electron beam energy of $\sim 100$ MeV~\cite{Suda2017} under the luminosity of $\sim 10^{27} \mathrm{cm}^{-2} \mathrm{s}^{-1}$, which is required for elastic electron scattering as mentioned above.  It should be stressed that this is not possible in a nucleus-nucleus collision even using $\sim$ GeV/u energetic RI beams.

\begin{figure}[htb]
    \centering
    \includegraphics[width=1.0\textwidth]{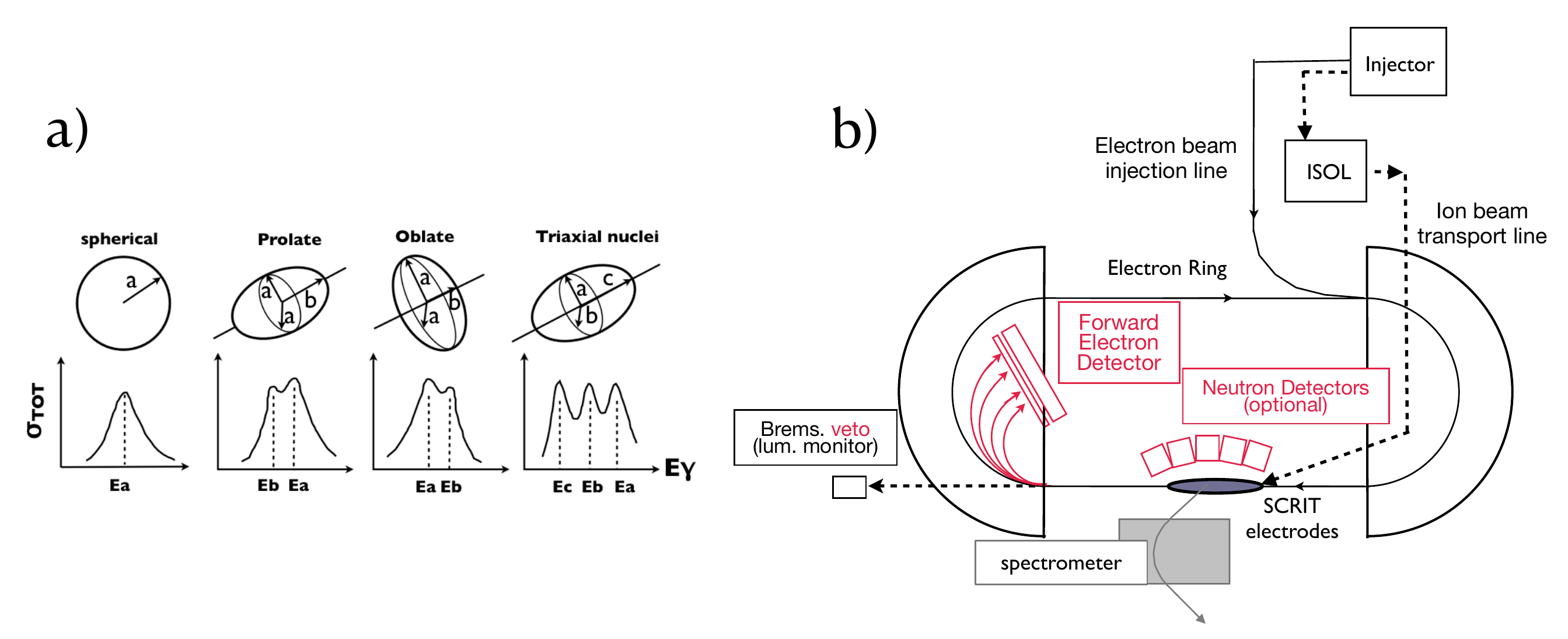}
    \caption{a) Photoabsorption cross section in the GDR region for nuclei having ground-state deformation.  b) A conceptual drawing of the future upgrade of SCRIT facility for photonuclear reaction measurements of radioactive nuclei~\cite{Suda2017}.  Those with the black lines indicate existing facilities and the red lines required equipment.}
    \label{fig:SCRITphotonuclear}
\end{figure}

One physics case-study at such a facility is shown in Fig.~\ref{fig:SCRITphotonuclear} a).  Photonuclear responses of exotic nuclei in the GDR region may reveal exotic deformations in the ground state including the triaxial deformation~\cite{Berger1977}.
This idea has been discussed as one of the possible future upgrades of the SCRIT facility as shown in Fig.~\ref{fig:SCRITphotonuclear} b).
Such studies could potentially be adapted to serve broader purposes, such as searches for permanent electric dipole moments 
in atomic nuclei, where detection of a nonzero result would imply sources of CP violation beyond the Standard Model.  The sensitivity of these searches would be enhanced by the appearance of pear-shaped deformation in the ground states of exotic nuclei~\cite{Gaffney2013,Butler:2019qox,Butler:2020rmc,Chishti2020,Behr:2022hym}. 
Radioactive beam studies at ISOLDE, e.g., have used Coulomb excitation~\cite{Gaffney2013,Butler:2020rmc} and other methods~\cite{Butler:2019qox,Chishti2020} to help identify the best-suited nuclei for such studies, and electron-scattering studies could potentially play an important complementary role. 

\subsection{Nuclear Astrophysics}

Radiative capture reactions, i.e. nuclear reactions in which the incident projectile is absorbed by the target nucleus and $\gamma$ radiation is then emitted, play a crucial role in nucleosynthesis processes in stars~\cite{Bru2015}, hence knowledge of their reaction rates at stellar energies is essential to understanding the abundance of the chemical elements in the universe. However, determination of these reaction rates has proven to be challenging, principally due to the Coulomb repulsion between initial-state nuclei and the weakness of the electromagnetic force. For example, the decay of unbound nuclear states by the emission of a particle of the same type as that captured, or by the emission of some other type of particle, is often 10$^3$–10$^6$ times more probable than decay by $\gamma$ emission.

The availability of an intense, low-energy electron beam rather than a photon beam allows the possibility to measure the inverse reaction. This idea has been previously proposed but not measured, and was more recently considered in detail~\cite{Fri2019} for the case of C$^{12}$($\alpha$,$\gamma$)$^{16}$O.  See contribution by {I.~Fri\v{s}\v{c}i\'{c}} in this proceedings.  In addition to a low-energy electron beam, a gas jet target, high resolution magnetic spectrometer and $\alpha$ recoil detector are required to carry out the experiment.

\section{QED Tests}
\subsection{Threshold Positron Production}

\begin{figure}[h]
  \centering
   \includegraphics[width=25pc]{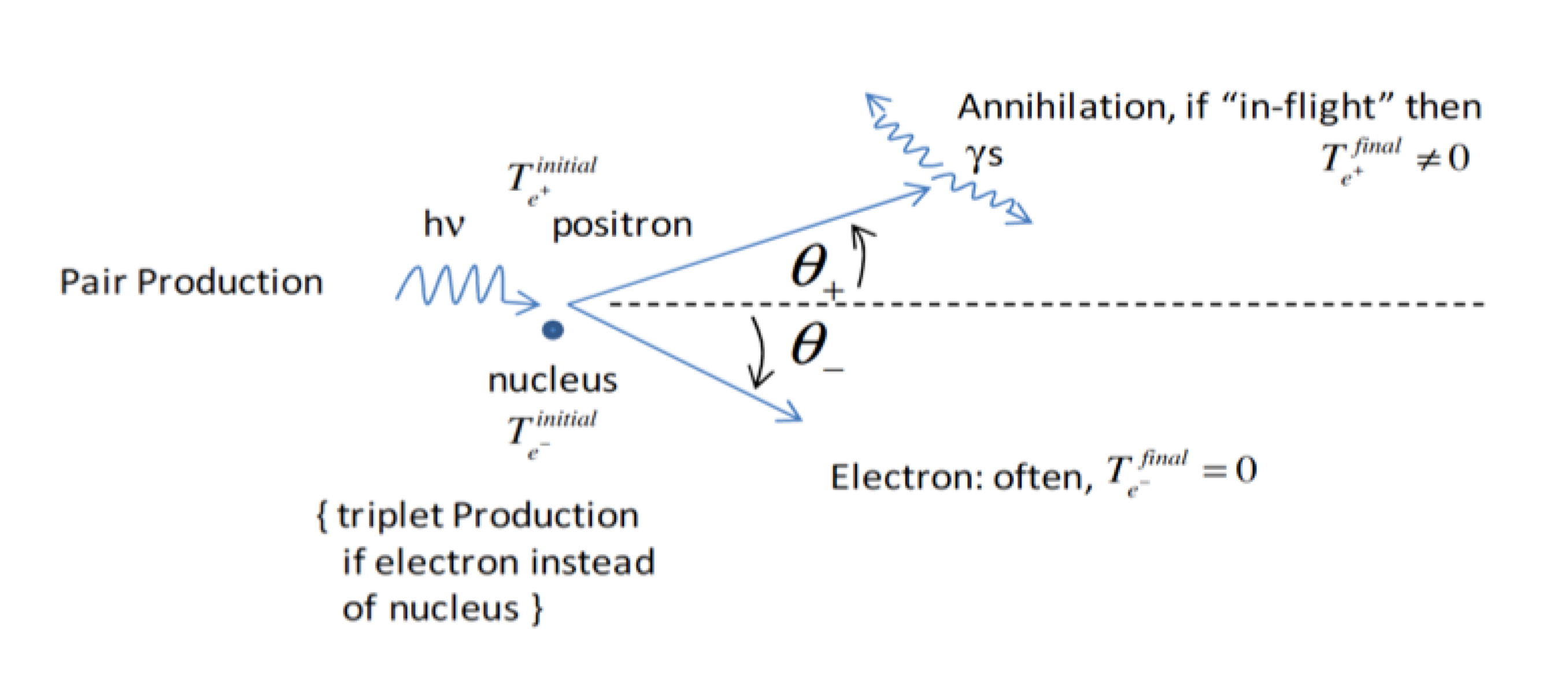}
    \caption{\label{Posprod}Pair and triplet $e^+e^-$ production from an atom.}
\end{figure}

In both pair and triplet production a positron and electron are produced spontaneously as a photon interacts with a strong electric field from either a nucleus (pair production) or an electron (triplet production) as shown in Fig.~\ref{Posprod}.

The threshold electron beam energy for these effects to take place is
\begin{equation}
(h \nu)_{\rm min} = 2 m_0 c^2 \left ( 1 + \frac{m_0}{M} \right ) \ ,
\end{equation}
where $m_0$ and $M$ are the incident electron and target masses, respectively and
\begin{eqnarray}
M &\equiv& M_{\rm nucleus} >> m_0\ ({\rm pair\ production})\\
M &=& m_0\ ({\rm triplet\ production}) \ . 
\end{eqnarray}
The mean kinetic energy given to each of the two particles is half of the available kinetic energy $T_{avail}$ (actually the positron gets a bit more energy because of the push from the positively charged nucleus).  The mean angle given to each of the two particles with respect to the incident electron directions is
\begin{equation}
    \overline{\theta}_{\pm} = \frac{2 m_0c^2}{T_{avail}} \ .
\end{equation}
The $1/T$ dependence is similar to bremsstrahlung: higher energy particles get more forward directed.  

\begin{figure}[h]
\includegraphics[width=22pc]{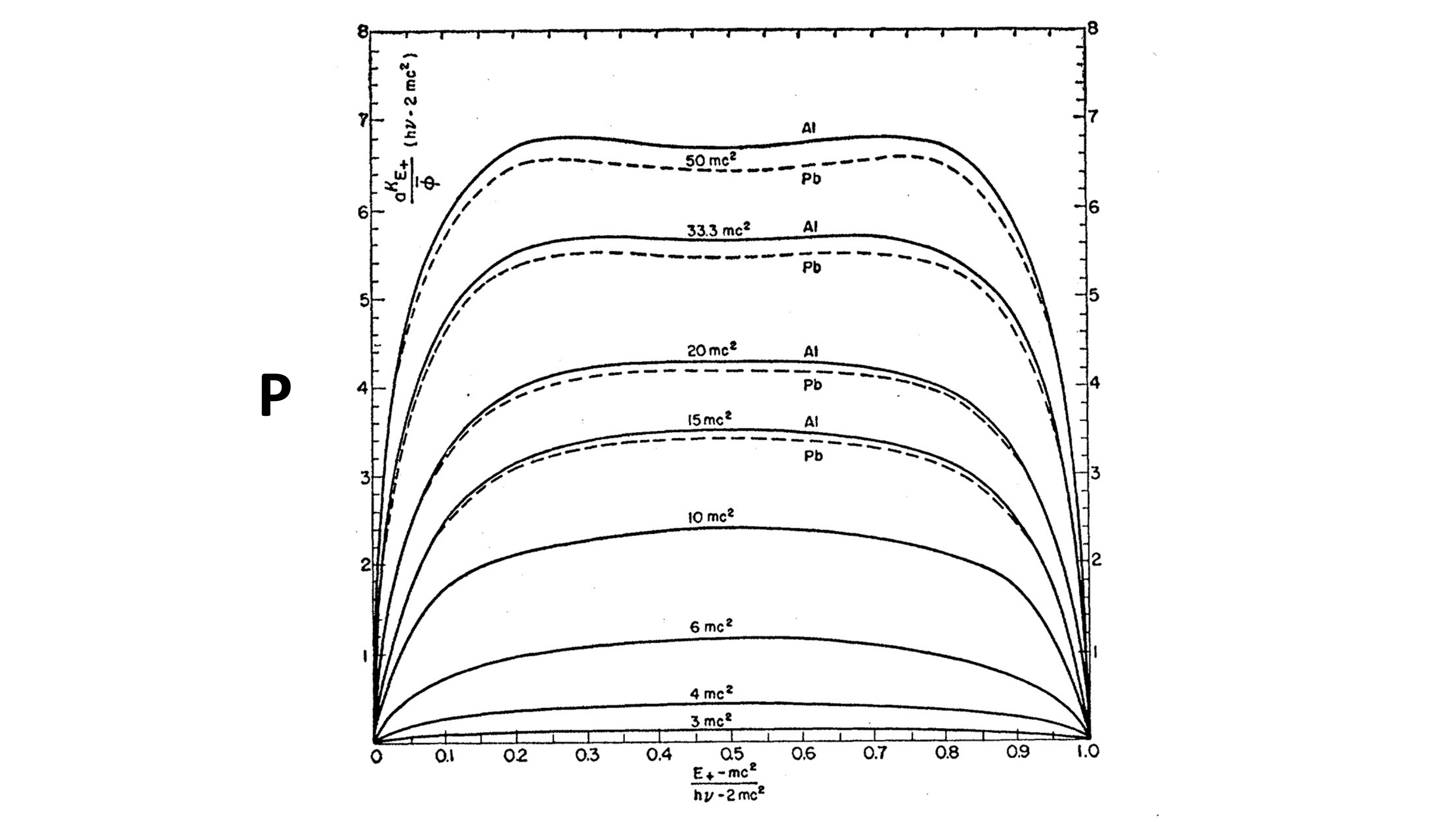}\hspace{2pc}%
\begin{minipage}[b]{14pc}\caption{\label{Posenergy}$P$ vs. kinetic energy fraction given to the positron from~\cite{Dav1952}. Notice the symmetry: energy not given to the positron is given to the electron and vice-versa.} 
\end{minipage}
\end{figure}

Bethe and Heitler~\cite{Bet1934} derived the cross section per atom for {\bf pair production} as
\begin{equation}
   \frac{d \kappa^{pair}_a}{d \overline{T}_{\pm}} = \sigma_0 \frac{Z^2}{T_{avail}} P \ ,
\end{equation}
where $P$ is a function shown in Fig.~\ref{Posenergy}, $Z$ is the nuclear charge and $\sigma_0 \equiv \frac{r^2_0}{137} = 4.8 \times 10^{-28} {\rm cm}^2 /{\rm atom}$.

In the case of {\bf triplet production}, the electric field is now from an electron, a very light particle, which becomes indistinguishable from the created particle.  The mean kinetic energy given to each of the three particles is now one-third of the available kinetic energy.  The threshold is now $4m_0c^2$.  The triplet production cross section is related to the pair production cross section as
\begin{equation}
    \kappa^{triplet}_a = \kappa^{pair}_a \cdot \frac{1}{CZ} \ ,
\end{equation}
where $C \approx 1$ and has no $Z$ dependence.

\begin{figure}[h]
\includegraphics[width=22pc]{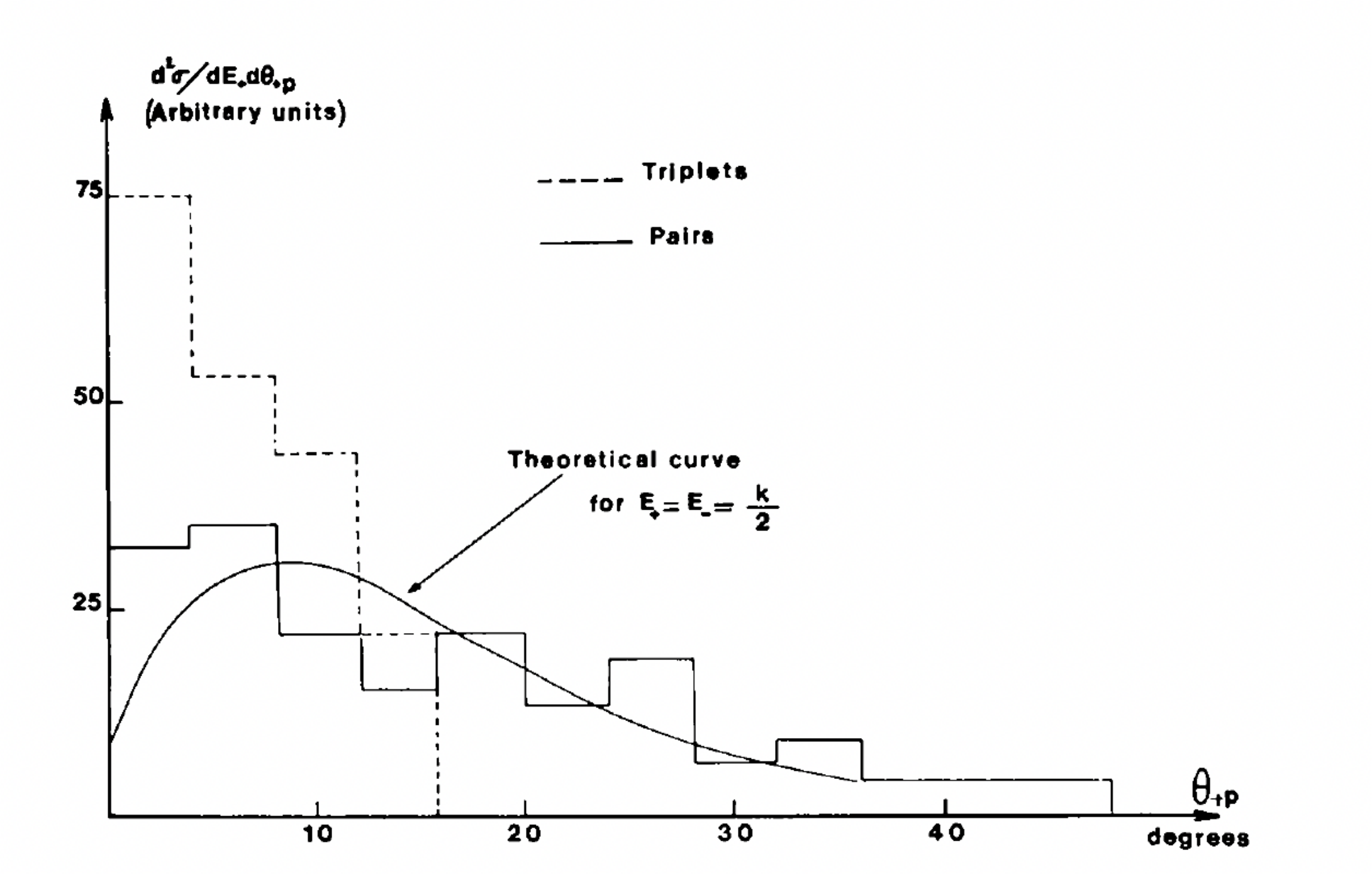}\hspace{2pc}%
\begin{minipage}[b]{14pc}\caption{\label{Tripdata} Positron angular distribution relative to the projected emission angle from~\cite{Roc1972}. Experimental histograms are given for triplet and pair production.  The theoretical curve is plotted for pair production.} 
\end{minipage}
\end{figure}

Measurements have been reported~\cite{Das1939,Shi1941,Ogl1945,Phi1949} with sporadic sightings of triplet events.
The first definitive observation of triplet production was in 1972~\cite{Roc1972} using a streamer chamber (Fig.~\ref{Tripdata}). At ARIEL, such an experiment could be mounted as well.  With 2~$\mu$A of 2.5~MeV electron beam on a 5 micron thick carbon foil (luminosity = $3 \times 10^{33}$ cm$^{-2}$ s$^{-1}$) the rate of positrons produced via the triplet mechanism is estimated (using~\cite{Roc1972}) at about 2 Hz into a spectrometer of solid angle 1 msr.
This should allow much more precise measurements of positrons at threshold than previously possible. Clearly, low electron beam energies of order 1-3 MeV are required. 

\subsection{Radiative M\o ller Scattering}

\begin{figure}[h]
  \centering
   \includegraphics[width=25pc]{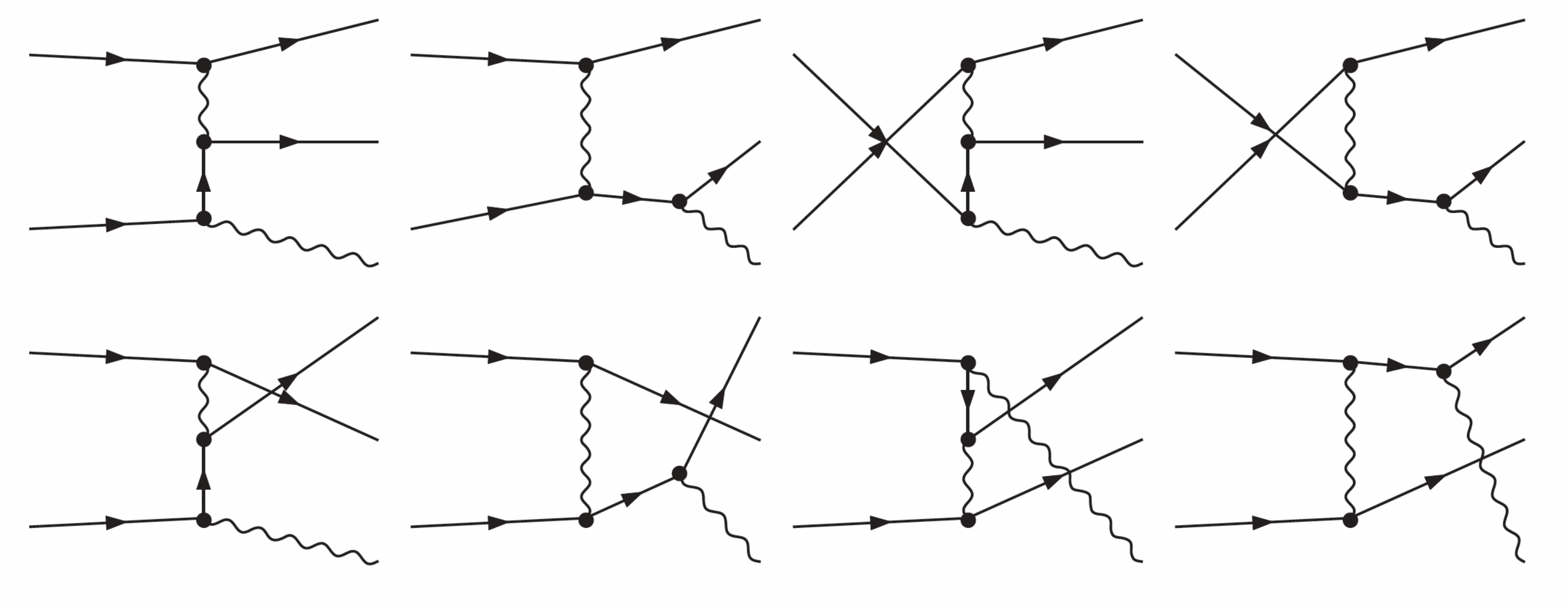}
    \caption{\label{FD}Feynman diagrams for radiative M\o ller scattering.}
\end{figure}

M\o ller (electron-electron) scattering is a background in electron scattering
experiments, and is a purely QED process at low energies. It is theoretically
straightforward to calculate, and has been so for decades. However, a modern
retrospective has revealed gaps in previous treatments, particularly the omission
of the electron mass in the calculation of the radiative diagrams shown in Fig.~\ref{FD}. Corrections due to these diagrams are typically included as a
multiplicative factor to the Born cross section:

\begin{equation}
    \frac{d \sigma}{d \Omega} \vert_{\rm soft} = (1 + \delta) \frac{d \sigma}{d \Omega} |_{\rm Born} \ ,
\end{equation}
with $\delta = \delta(\Delta E,\Omega)$. This traditional method requires defining a cut-off $\Delta E$,
the maximum amount of energy a photon can carry away for which the event
passes acceptance cuts. For an experiment having spectrometers with small,
well-defined energy and angular acceptances, this formulation of the radiative
corrections can be applied straightforwardly.

Radiative corrections for hard-photon bremsstrahlung emission in both M\o ller and Bhabha scattering have been performed~\cite{Eps2016} in a consistent approach without ultra-relativistic approximations and permit  a complete analysis at next-to-leading-order.
\begin{figure}[h]
\includegraphics[width=14pc]{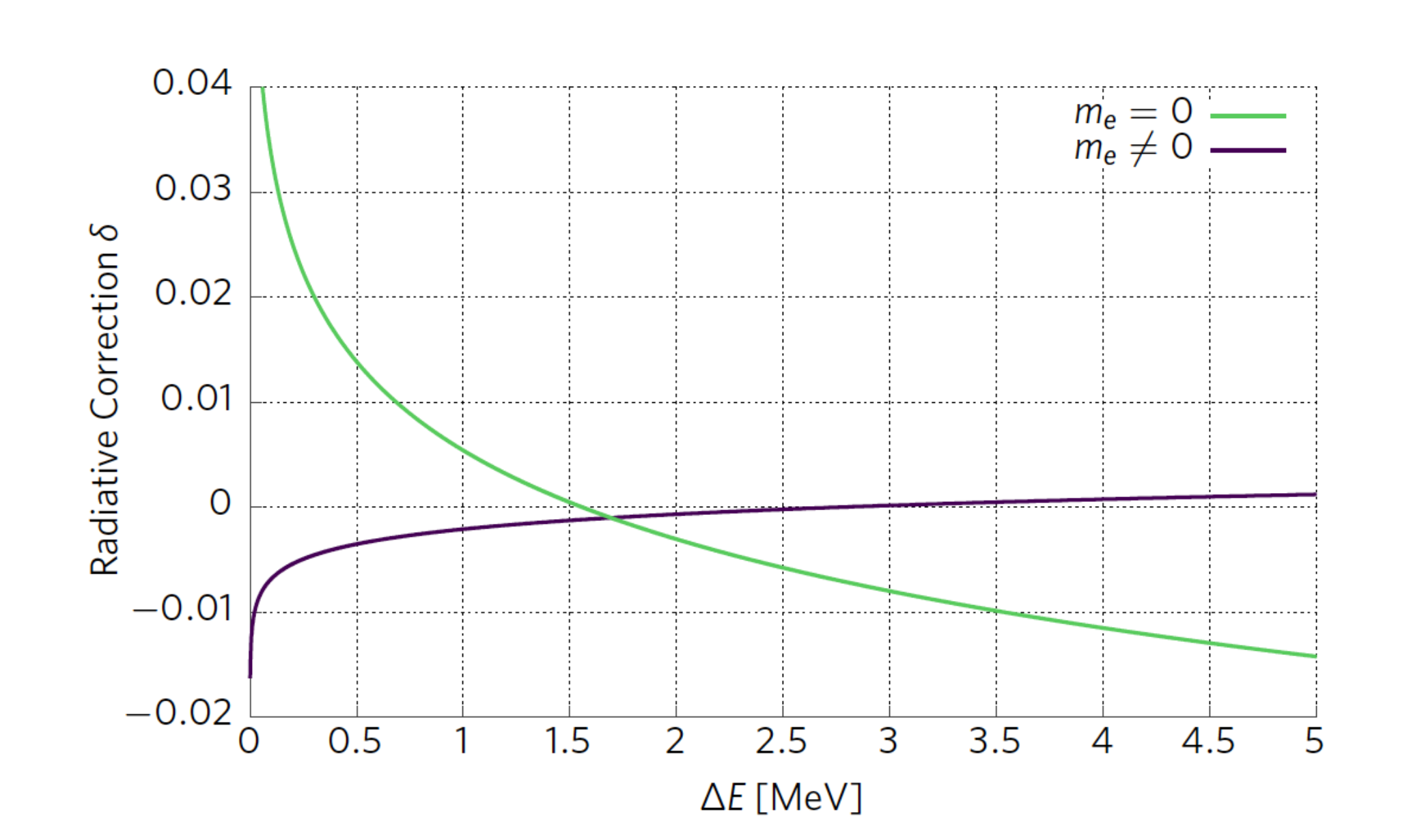}\hspace{2pc}%
\begin{minipage}[b]{14pc}\caption{\label{label}Comparison of the M\o ller radiative correction term $\delta$ for a 100 MeV electron beam at 5$^\circ$ in the CM frame for different electron masses from~\cite{Eps2016}.}
\end{minipage}
\end{figure}

Measurement of radiative M\o ller scattering as a function of energy between 10 and 100 MeV is highly desirable to validate the correction procedure.  This has been carried out at 2.5 MeV (where radiative effects are negligible) using a focusing spectrometer~\cite{Eps2020} by the DarkLight collaboration.  The desired measurements can be straightforwardly carried out using the ARIEL electron linac with the apparatus proposed for the DarkLight experiment.

\section{Searches for Dark Matter and BSM Physics}

A dark sector search is already being undertaken at the ARIEL e-linac by the DarkLight collaboration, so it is natural to explore related searches that might be done with the same accelerator. The physics scope for the ARIEL e-linac is limited by its maximum energy of 50 MeV, and it is important to note that the low-energy dark sector landscape in both visible and invisible final states is well populated by a range of existing experiments. Finding a new area where an experiment at TRIUMF can make a strong contribution depends on either identifying an uncovered niche outside of the benchmark simplified models typically used by the field, as DarkLight did with protophobic bosons~\cite{Feng:2016jff}, or by 
taking advantage of the ARIEL e-linac's unique strength: the high intensity of its beam. The clearest opportunities presented are, e.g. rare processes at low mass which are too time-consuming to probe with lower intensity beams, or for which the low energy allows for some particular access to challenging kinematics or limits SM backgrounds. 
In what follows we offer concrete examples that illustrate the possibilities. 

\subsection {Delbr\"{u}ck Scattering to Search for Dark Photons of sub-MeV in Mass}

A dark photon, the boson $A'$ associated with a U(1) gauge symmetry of a dark sector~\cite{Holdom:1985ag}, can mix with the 
SM photon with strength $\varepsilon$, with $\varepsilon \ll 1$, to produce millicharged couplings, namely, 
${\cal L}_{\rm mix} = \varepsilon A^{'}_{\mu} J_{\rm EM}^{\mu}$, with the electromagnetically charged fermions of the SM. The nature of this ``portal'' interaction between the dark and visible sectors allows an enormous range of dark photon masses and mixing parameters to appear, and
we must look to experiment to constrain the possibilities. Studies in $e^-e^+$ collisions~\cite{BaBar:2014zli,BaBar:2017tiz}
have been very effective in limiting the parameter space for dark photons in excess of $2m_e$ in mass, 
and there are a broad sweep of astrophysical and cosmological constraints as well~\cite{Caputo:2021eaa,Antypas:2022asj}. 
However, if the dark photon mass $m_{A'}$ satisfies $m_{A'}\lesssim 2m_e$, then 
its mixing with the photon 
guarantees that the decay $A'\to 3\gamma$ can occur, though the dark photon in this case can be quite long-lived, 
because the decay is a one-loop process with a rate that scales as $\varepsilon^2 \alpha^4$. 

Nevertheless, there may be 
discoverable phase space yet 
to explore~\cite{McDermott:2017qcg}. 
This conclusion largely stems from a proper assessment of the decay rate: 
Earlier work computed the decay rate 
in the Euler-Heisenberg limit~\cite{Pospelov:2008jk}, which assumes that $m_{A'} \ll m_e$, but 
the exact one-loop calculation yields a result some 10-100 times larger if 
$850\, {\rm keV} \lesssim m_{A'} \lesssim 1\, {\rm MeV}$,  
shortening the dark photon lifetime and altering the
excluded phase space significantly~\cite{McDermott:2017qcg}. 
Referring to Fig.~4 of \cite{McDermott:2017qcg}, we note that $850\, {\rm keV} \lesssim m_{A'} \lesssim 1\, {\rm MeV}$ and $10^{-5} \lesssim \varepsilon' \lesssim 10^{-4}$ represents an interesting window of opportunity~\cite{McDermott:2017qcg}, with constraints coming from big-bang nucleosynthesis, which would appear to demand $\tau_{A'} < 1 \, \rm sec$, and from the anomalous magnetic moment of the electron, $(g-2)_e$~\cite{Hanneke:2008tm,Pospelov_secluded_PhysRevD.80.095002}. 
A constraint also comes from nonobservation of $A^\prime + e \to e + \gamma$ in the LSND 
experiment, presuming the production of $A^\prime$ from $\pi^0 \to A^\prime \gamma$ decay~\cite{Pospelov_light_2017kep}. 
The authors of \cite{Pospelov_light_2017kep} thus contend this region is excluded, but the LSND constraint can be evaded
through a non-minimal dark vector model, such as the protophobic gauge boson. 
We note, too, the experimental controversy in the precise value of the fine-structure constant $\alpha$ from atom interferometry~\cite{Parker:2018vye,Morel:2020dww}, noting \cite{Gardner:2019mcl} for context, with resolutions that would alter the $(g-2)_e$ constraint of \cite{Pospelov_secluded_PhysRevD.80.095002} somewhat. 

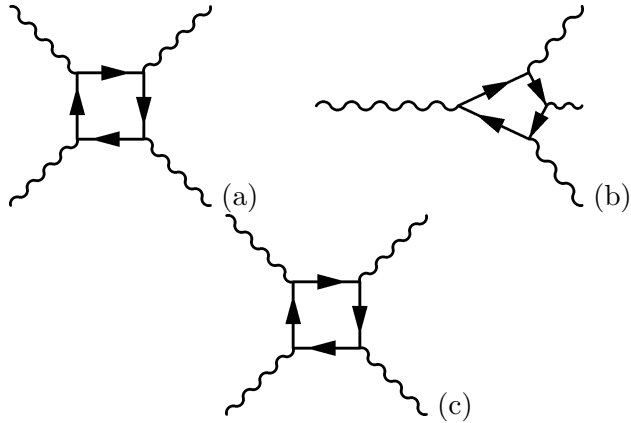
\begin{figure}
\begin{center}
\begin{fmffile}{diagram}
\begin{fmfgraph*}(75,75)
    \fmfstraight
    \fmfleft{i1,i2}
    \fmfright{o1,o2}
    \fmf{photon,label=$\gamma^*$}{i1,w1}
    \fmf{photon,label=$\gamma^*$}{w2,o1}
    \fmf{fermion}{w2,w1}
     \fmf{fermion}{w3,w4}
     \fmf{fermion}{w4,w2}
     \fmf{fermion}{w1,w3}
     \fmf{photon,label=$\gamma$}{i2,w3}
     \fmf{photon,label=$\gamma$}{w4,o2}
      \fmffreeze
\end{fmfgraph*} 
(a)
\hspace{0.5cm}
\begin{fmfgraph*}(100,75)
    \fmfstraight
    \fmfleft{i1}
    \fmfright{o1,o2,o3}
    \fmf{photon,label=$A^\prime$}{i1,w1}
    \fmf{photon,label=$\gamma$}{w2,o1}
    \fmf{fermion}{w2,w1}
     \fmf{fermion}{w4,w3}
     \fmf{fermion}{w3,w2}
     \fmf{fermion}{w1,w4}
     \fmf{photon,label=$\gamma$}{w4,o3}
     \fmf{photon,label=$\gamma$}{w3,o2}
      \fmffreeze
\end{fmfgraph*} 
(b)
\hspace{0.75cm}

\begin{fmfgraph*}(75,75)
    \fmfleft{i1,i2}
    \fmfright{o1,o2}
    \fmf{photon,label=$\gamma^*$}{i1,w1}
    \fmf{photon,label=$\gamma^*$}{w2,o1}
    \fmf{fermion}{w2,w1}
     \fmf{fermion}{w3,w4}
     \fmf{fermion}{w4,w2}
     \fmf{fermion}{w1,w3}
     \fmf{photon,label=$\gamma$}{i2,w3}
     \fmf{photon,label=$A^\prime$}{w4,o2}
     \fmffreeze
\end{fmfgraph*} 
(c)
\end{fmffile} 

\end{center}
\caption{
Representative Feynman diagrams to illustrate (a) Delbr\"uck scattering, which is 
light scattering in the Coulomb field of an atom or nucleus, as indicated by the appearance of $\gamma^*$, 
(b) $A'\to 3\gamma$ decay, and (c) Delbr\"uck scattering modified to probe a sub-MeV dark photon, with the appearance of 
missing momentum giving the experimental signature. Note that crossed diagrams exist but have not been shown. 
The solid lines denote any of the electrically charged fermions of the SM, with the electron/positron giving the 
dominant contribution. 
}
\label{Fig:darkphotondecay}
\end{figure}

Delbr\"uck scattering, in which a photon is deflected in a strong Coulomb field due to vacuum polarization, would be sensitive to the $A^\prime$ in this region without assumption concerning the production mechanism.
Pertinent diagrams are illustrated in Fig.~\ref{Fig:darkphotondecay}. We note that at ARIEL energies, Delbr\"uck scattering has been clearly identified~\cite{Milstein:1994zz}. The suggested new-physics search would require the ability to assess the existence of missing momentum or energy.

\subsection{$|\Delta B| = 2$ Processes in Nuclei}
Experimental limits on processes that would change baryon number by one unit, i.e., $|\Delta B| =1$, are among
the most stringent known to science~\cite{Workman:2022ynf,Berryman:2022zic}.  Experimental studies on processes that would change baryon number by two or more units, however, have been much more limited~\cite{FileviezPerez:2022ypk}, although their physical origin can be altogether different~\cite{Mohapatra:1980qe,Arnold:2012sd,Gardner:2018azu}. 
There is clear reason to search for such effects:  The existence of $|\Delta (B-L)|=2$ violation is necessary to make a massive neutrino its own antiparticle~\cite{Weinberg:1979sa}, and observable $B$ and/or $B-L$ violation in the quark sector would help to identify its dynamical mechanism.

The ARIEL accelerator facility's intense electron beam would be well suited to the search for $|\Delta B| =2$ through low-energy channels.  Of particular note are the processes $e^- p \to e^+ \bar p$ or $e^- p \to {\bar n} \bar\nu$~\cite{Gardner:2018azu}, with the $B-L\,$-violating channel
$e^- n \to e^- \bar n$ also open to study~\cite{Gardner:2017szu} --- here the use of a deuteron target could prove advantageous~\cite{Gardner:2017szu,Gardner:2018azu}. The estimated event rates for the $B$ violating, but $B-L$ conserving, processes we have noted are no more than ${\cal O}(10)$ events per year~\cite{GYprep,SVGproc}.
Thus the low energy of the existing electron linac at ARIEL is a key advantage, rather than a limitation, both in controlling backgrounds and in detecting the produced anti-nucleons.  At ARIEL, backgrounds are controlled by the extremely low energy of the electron scattering process; in particular, the accelerator operates at energies far below pion production threshold. 
Moreover, the produced anti-nucleon would be at sufficiently low energy that the annihilation should be prompt and thus occur within the target, typically yielding a five-pion final state, as determined in R\&D studies of searches for $n-{\bar n}$ oscillations~\cite{Phillips:2014fgb,Addazi:2020nlz}. In the ARIEL environment, the supposed five-pion final state signal should be quite striking. Their total charge, in a proton target,  should signal the electric charge of the anti-nucleon and hence the precise process. This search could potentially run parasitically at ARIEL.

\subsection{Beam-Dump Experiments to Search for Dark Matter}

The high intensity of the ARIEL e-linac makes it an interesting potential location to consider a beam dump dark matter experiment as well. The sensitivity relative to DarkMESA, LDMX, and other light DM searches should be examined using an appropriate benchmark model and the usefulness of such an experiment determined in light of the fact that it would necessarily begin much later than already-approved experiments.

\begin{figure}[!htb]
    \centering
    \includegraphics[width=25pc]{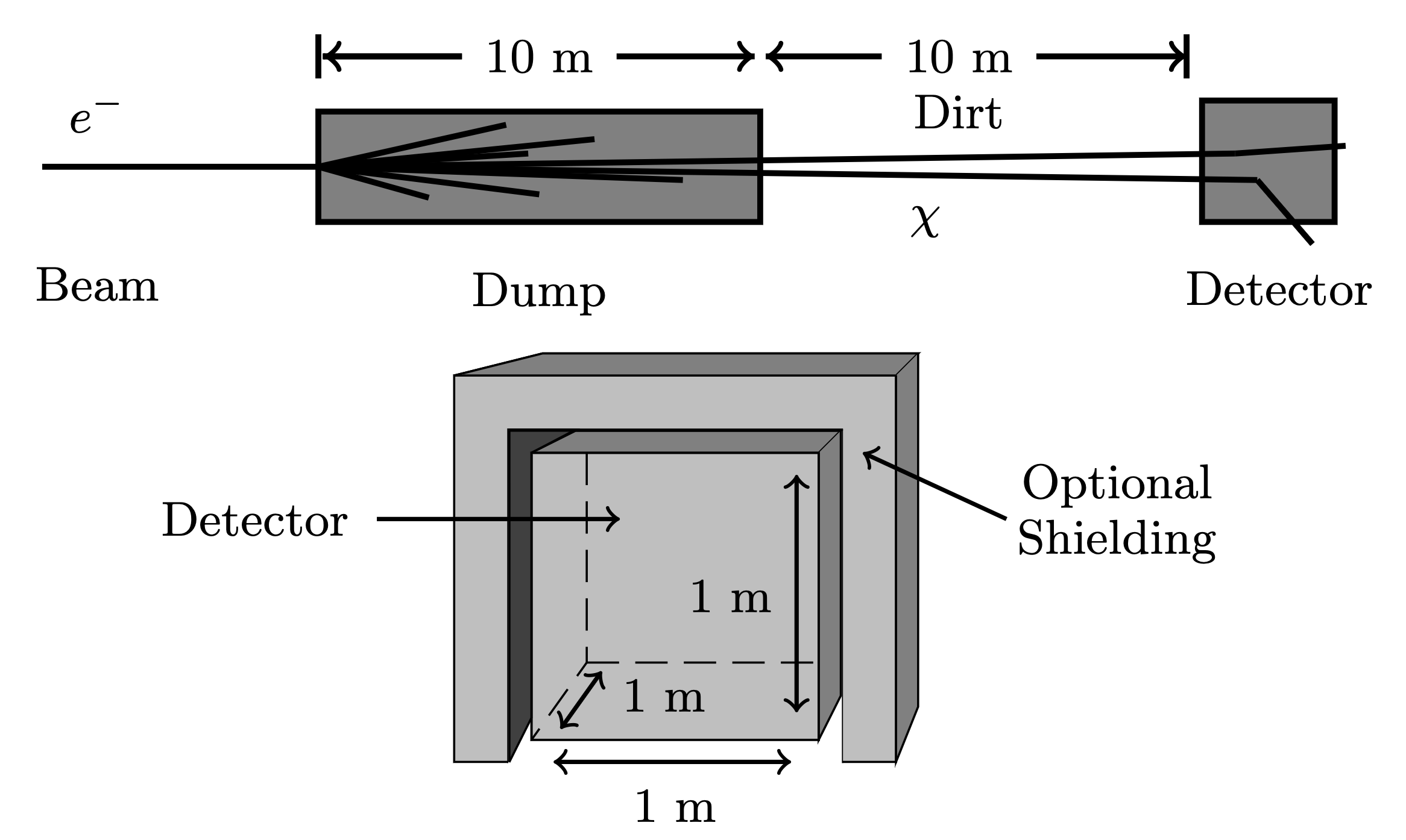}
    \caption{Schematic of a beam dump experiment searching for light dark matter, from~\cite{Izaguirre:2013uxa}.}
    \label{fig:beamdump}
\end{figure}

Dark matter searches at beam dumps are relatively model-independent. A high-current beam is directed to a target where dark sector particles may be created. Enough shielding to stop essentially all Standard Model particles is placed behind or incorporated into the target, and a detector behind the shielding searches for dark matter particles exiting the dump. The detector is sensitive to dark matter particles via their scattering in the target material. Examples of such experiments, interpretations, and proposals include BDX~\cite{BDX:2017jub}, DarkMESA (see contribution by L. Doria in this proceedings), E137~\cite{Batell:2014mga}, the SHiP Scattering Neutrino Detector~\cite{SHiP:2020noy}, COHERENT~\cite{COHERENT:2019kwz}, and MiniBooNE ( in its dedicated dark matter configuration~\cite{MiniBooNEDM:2018cxm}). A schematic of a beam dump DM experimental setup, including a simplified detector of moderate scale, is shown in Fig.~\ref{fig:beamdump}.

Beam-dump experiments probe the same types of final states as missing-momentum or missing-energy based experiments (LDMX, NA64). They share with them the advantage that these approaches, unlike visible final state searches, are relatively agnostic to the lifetime of the mediator particle(s) and to the decay chain producing the dark matter and can therefore constrain a wide range of models~\cite{Berlin:2018bsc}. Scattering-based beam dump experiments are, however, disadvantaged relative to missing-momentum experiments in that they depend on an additional interaction and therefore gain a $\sim \alpha_D \epsilon^2$ suppression factor~\cite{LDMX:2018cma}. The difference could be made up with sufficient beam intensity, since missing momentum experiments must run at low current, and the ARIEL e-linac's uniquely high power makes it worth investigating as a potential site.

A viable location for a beam dump experiment does exist at the ARIEL e-linac. A small room behind the ARIEL targets could host a detector, allowing this new experiment to take data parasitically to ARIEL and without disrupting the operation of DarkLight or another future experiment in its position. One potential experimental challenge is the low momentum transfer from scattering DM particles in the detector at the mass and momentum ranges accessible at ARIEL. This is not insurmountable but would require thoughtful detector design.

\begin{figure}[!htb]
    \centering
    \includegraphics[width=0.65\textwidth]{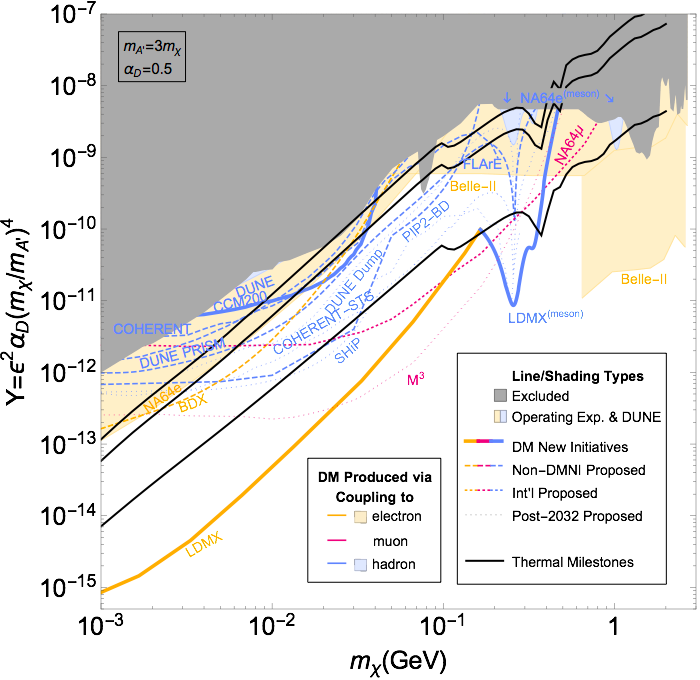}
    \caption{Constraints on light dark matter in the context of invisible decays of dark photons from current, future, and proposed experiments~\cite{Snowmass-DP}.}
    \label{fig:dp_inv}
\end{figure}

\begin{figure}[!htb]
    \centering
    \includegraphics[width=0.8\textwidth]{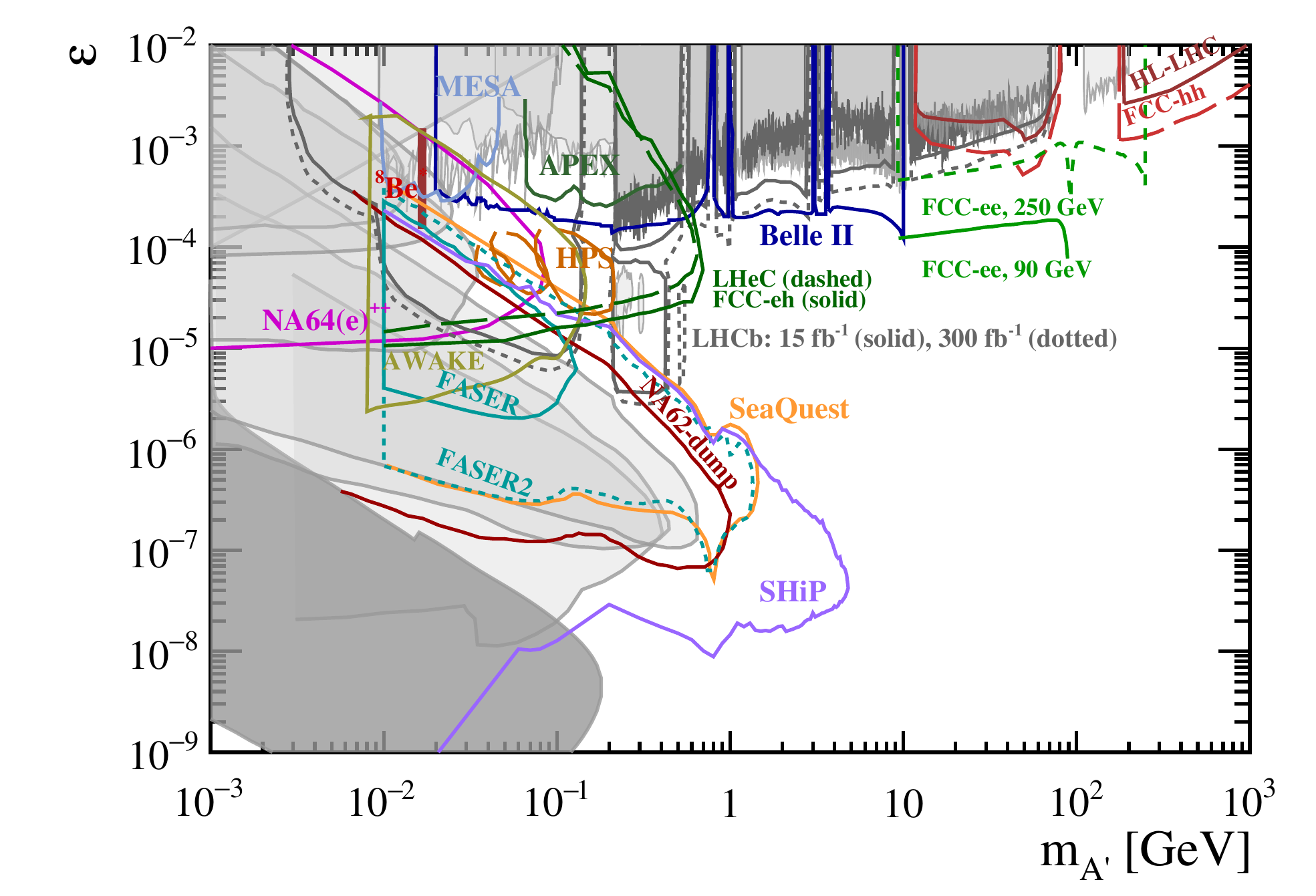}
    \caption{Constraints on dark photons detectable through visible decays in the context of current, future, and proposed experiments.  Grayed regions represent existing limits, in the assumption of flavor-independent couplings.  Colored lines represent future projected exclusions~\cite{Fabbrichesi_2021}.}
    \label{fig:dp_vis}
\end{figure}

The current and proposed future exclusion landscape for light dark matter in the context of invisible dark photon decays is shown in Fig.~\ref{fig:dp_inv}. The dominant sensitivity in the sub-10~MeV mass range is projected to come from LDMX. 
For reference, the landscape for visible dark photon decays is shown in Fig.~\ref{fig:dp_vis}.

It could also be considered whether there is any scope for a missing-energy or missing-momentum search at the ARIEL e-linac. The practical advantages of a classic beam dump experiment at TRIUMF would be negated, since such an experiment could not operate parasitically behind the ARIEL targets, but it would offer the opportunity for significantly higher sensitivity. This approach would require a solution to the experimental challenges of missing momentum reconstruction in the presence of extremely high electron multiplicity in the beam bunches. 

\begin{figure}[!htb]
    \centering
    \includegraphics[width=0.8\textwidth]{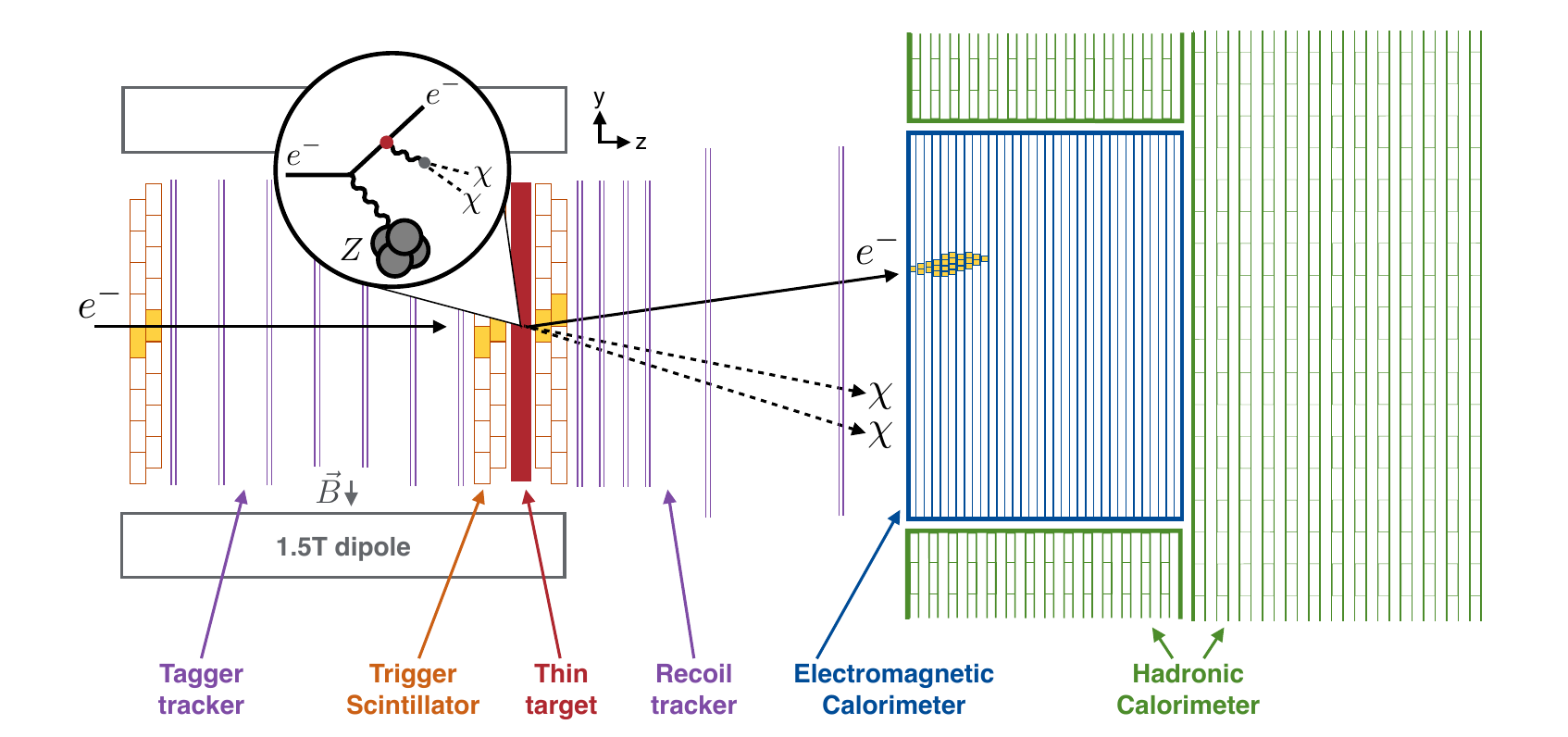}
    \caption{Illustration of the production of dark matter
    particles $\chi$ through dark photon decay, apropos to 
            the LDMX detector, from \cite{Akesson:2022vza}.
            Replacing $\chi\chi$ with 
            $\chi \bar \chi$ illustrates the possibility 
            of a dark gauge boson coupled to baryon number, 
            as discussed in the text.}
    \label{fig:LDMX}
\end{figure}

We conclude this section by noting another way in which dark-sector searches at the ARIEL facility could be complementary to accelerator-based DM experiments planned elsewhere. At the proposed LDMX experiment~\cite{Berlin:2018bsc,LDMX:2018cma,Akesson:2022vza}, e.g., invisible decays of a dark photon $A'$ can be probed through its missing-momentum signature. That is, as shown in Fig.~\ref{fig:LDMX}, an $A'$ produced through bremsstrahlung in electron-nucleus scattering can decay invisibly via $A'\to \chi\chi$, and this possibility can be probed through the measurement of the momentum of the final-state electron. This may be a unique possibility to probe $\chi$ because $\chi$ carries no SM harges. 
However, in place of the usual dark photon, the gauge boson mediator could couple to baryon number instead~\cite{Nelson:1989fx,Tulin:2014tya,Berryman:2021jjt,Berryman:2022zic}.The gauge boson in such a model couples to neutrons and photons, and it can couple to electrically charged particles, through kinetic mixing with the SM photon. In this case the dark gauge boson can still be produced via bremsstrahlung, but it could decay to a $\chi \bar \chi$ pair, in which $\chi$ carries baryon number $B=1/3$. Figure~\ref{fig:LDMX} also illustrates this possibility upon the replacement of $\chi\chi$ with $\chi\bar\chi$.

Although this model would generate a missing momentum signature, $\chi$ and $\bar \chi$ could also interact with neutrons and protons in the far detector. One possibility with a striking observable signature concerns {\it destabilization} of the nucleus.  Here the incoming DM particle $\chi$ stimulates the decay of a nucleus in the far detector via the neutron-decay process $\bar \chi n \to \chi \chi$.  We note that $n\to 3 \chi$ has also been considered in the context of the neutron lifetime anomaly~\cite{Strumia:2021ybk}, noting \cite{Fornal:2018eol}, and the decay rate need not be exceedingly small~\cite{Berryman:2022zic}. 
The breakup of the nucleus, with roughly a GeV of energy loss, could be documented through gamma detectors placed at the far detector. The latter could be sufficiently removed from the electron beam environment to make such a photon fingerprinting process possible.  The viability of such an approach has been studied in work by KamLAND~\cite{KamLAND:2005pen}, and references therein. 

\subsection{Search for X17 in $e^+e^-$ Decay of Giant Dipole Resonance}
It has been suggested that the X17 may also be able to be produced in the decay of giant nuclear resonances.  Both Isoscalar Giant Monopole Resonances (ISGMR) and Isovector Giant Dipole Resonances (IVGDR) can be readily excited in an electron beam fixed target experiment, and the typical energy scale of these resonances (ISGMR) follows Eq.~\ref{eqn:isgmr_scaling}.  For atomic numbers above about 20, these can be readily reached with a 30-50~MeV beam.
\begin{equation}
    \label{eqn:isgmr_scaling}
    E_c=80A^{-1/3} \mathrm{MeV}
\end{equation}
Final states may carry angular distributions associated with the excited multipole but should be randomly oriented in the detector frame, producing a signature that is independent of the angle with respect to the electron beam, distinguishing those decay particles from the forward-peaking prompt events.  

Unfortunately, electron scattering has a large contribution from the radiative tail, which increases the background associated with identifying a resonance, and identifying the particles from that resonance's decay, substantially.  At the collision energies of the current or upgraded ARIEL accelerator, techniques to separate prompt and resonant final states spatially or temporally would need to have exquisite resolutions that are not currently feasible.  The lifetime of the resonance is extremely short (typically $10^{-21}$ to $10^{-19}$ s), and its recoil energy essentially zero.  At higher beam energies these constraints are relaxed, but these are beyond the upgrades currently discussed for ARIEL.

We note that different excitation techniques, in particular tagged photons, may make giant resonances a more fruitful area both for exotic particle searches and other studies of resonance structures, but would require the development of a photon-tagging station on the beamline.

\section{Positron Source for Materials Science Research}

Low energy positrons are a unique probe of materials and surfaces that allow measurements of surface properties not possible with other probes such as electrons or X-rays.  Though nuclear decay sources are available for positron studies, accelerator sources can provide higher fluxes and much higher brightness.  See~\cite{Benson} for a detailed consideration of such a source at the Low Energy Recirculator Facility at Jefferson Laboratory, Newport News, VA, USA.

ARIEL upgraded in energy and operating with energy recovery has the potential to produce a positron source of great interest for materials science research.  The addition of spin-polarization capability to the positron source would enhance the research scope.  Full realization of such a capability would likely attract a high demand from users.

\bigskip
\section{Summary}
\bigskip

A high-intensity electron beam at low energies as can be provided at ARIEL offers the potential to pursue a suite of important scientific questions.  We have presented many here, along with initial exploration of the challenges and potential reach for each.  Several require significant investment to realize more sophisticated capabilities.  Further study will be essential to determine whether the investment is merited.     

\section*{References}
\medskip

\bibliography{main.bbl}

\begin{thebibliography}{10}

\bibitem{pohlSizeProton2010}
Randolf Pohl, Aldo Antognini, Francois Nez, Fernando~D. Amaro, Francois
  Biraben, Joao M.~R. Cardoso, Daniel~S. Covita, Andreas Dax, Satish Dhawan,
  Luis M.~P. Fernandes, Adolf Giesen, Thomas Graf, Theodor~W. Hansch, Paul
  Indelicato, Lucile Julien, Cheng-Yang Kao, Paul Knowles, Eric-Olivier
  Le~Bigot, Yi-Wei Liu, Jose A.~M. Lopes, Livia Ludhova, Cristina M.~B.
  Monteiro, Francoise Mulhauser, Tobias Nebel, Paul Rabinowitz, Joaquim M.~F.
  {dos Santos}, Lukas~A. Schaller, Karsten Schuhmann, Catherine Schwob, David
  Taqqu, Joao F. C.~A. Velos, and Franz Kottmann.
\newblock The size of the proton.
\newblock {\em Nature}, 466:213--216, 2010.

\bibitem{bernauerHighprecisionDeterminationElectric2010}
J.~C. Bernauer, P.~Achenbach, C.~Ayerbe Gayoso, R.~Bohm, D.~Bosnar,
  L.~Debenjak, M.~O. Distler, L.~Doria, A.~Esser, H.~Fonvieille, J.~M.
  Friedrich, J.~Friedrich, M.~{Gomez Rodriguez de la Paz}, M.~Makek, H.~Merkel,
  D.~G. Middleton, U.~Muller, L.~Nungesser, J.~Pochodzalla, M.~Potokar,
  S.~Sanchez~Majos, B.~S. Schlimme, S.~Sirca, Th~Walcher, and M.~Weinriefer.
\newblock High-precision determination of the electric and magnetic form
  factors of the proton.
\newblock {\em Phys.Rev.Lett.}, 105:242001, 2010.

\bibitem{mohrCODATARecommendedValues2008}
Peter~J. Mohr, Barry~N. Taylor, and David~B. Newell.
\newblock {{CODATA}} recommended values of the fundamental physical constants:
  2006.
\newblock {\em Rev. Mod. Phys.}, 80(2):633--730, 06 2008.

\bibitem{Xiong:2019umf}
W.~Xiong et~al.
\newblock {A small proton charge radius from an electron\textendash{}proton
  scattering experiment}.
\newblock {\em Nature}, 575(7781):147--150, 2019.

\bibitem{arrington2007}
J.~Arrington, W.~Melnitchouk, and J.~A. Tjon.
\newblock Global analysis of proton elastic form factor data with two-photon
  exchange corrections.
\newblock {\em Physical Review C}, 76(3):035205, 09 2007.

\bibitem{Alarcon:2018zbz}
J.~M. Alarc\'on, D.~W. Higinbotham, C.~Weiss, and Z.~Ye.
\newblock {Proton charge radius extraction from electron scattering data using
  dispersively improved chiral effective field theory}.
\newblock {\em Phys. Rev. C}, 99(4):044303, 2019.

\bibitem{PhysRevC.90.015206}
J.~C. Bernauer, M.~O. Distler, J.~Friedrich, Th. Walcher, P.~Achenbach,
  C.~Ayerbe~Gayoso, R.~B\"ohm, D.~Bosnar, L.~Debenjak, L.~Doria, A.~Esser,
  H.~Fonvieille, M.~G\'omez Rodr\'{\i}guez de~la Paz, J.~M. Friedrich,
  M.~Makek, H.~Merkel, D.~G. Middleton, U.~M\"uller, L.~Nungesser,
  J.~Pochodzalla, M.~Potokar, S.~S\'anchez~Majos, B.~S. Schlimme,
  S.~\ifmmode~\check{S}\else \v{S}\fi{}irca, and M.~Weinriefer.
\newblock Electric and magnetic form factors of the proton.
\newblock {\em Phys. Rev. C}, 90:015206, 07 2014.

\bibitem{Schlimme:2021gjx}
B.~S. Schlimme et~al.
\newblock {Operation and characterization of a windowless gas jet target in
  high-intensity electron beams}.
\newblock {\em Nucl. Instrum. Meth. A}, 1013:165668, 2021.

\bibitem{Antognini:2021icf}
Aldo Antognini, Franz Kottmann, and Randolf Pohl.
\newblock {Laser spectroscopy of light muonic atoms and the nuclear charge
  radii}.
\newblock {\em SciPost Phys. Proc.}, 5:021, 2021.

\bibitem{Katayama2003}
T.~Katayama, T.~Suda, and I.~Tanihata.
\newblock {Status of MUSES project and electron RI collider at RIKEN}.
\newblock In {\em Physica Scripta T}, volume 104, 2003.

\bibitem{Wakasugi2004}
M.~Wakasugi, T.~Suda, and Y.~Yano.
\newblock {A new method for electron-scattering experiments using a
  self-confining radioactive ion target in an electron storage ring}.
\newblock {\em Nuclear Instruments and Methods}, 532(1-2), 2004.

\bibitem{Tsukada2017}
K.~Tsukada, A.~Enokizono, T.~Ohnishi, K.~Adachi, T.~Fujita, M.~Hara, M.~Hori,
  T.~Hori, S.~Ichikawa, K.~Kurita, K.~Matsuda, T.~Suda, T.~Tamae, M.~Togasaki,
  M.~Wakasugi, M.~Watanabe, and K.~Yamada.
\newblock {First Elastic Electron Scattering from Xe 132 at the SCRIT
  Facility}.
\newblock {\em Physical Review Letters}, 118(26):1--5, 2017.

\bibitem{Suda2017}
Toshimi Suda and Haik Simon.
\newblock {Prospects for electron scattering on unstable, exotic nuclei}.
\newblock {\em Progress of Particle and Nuclear Physics}, 96:1--31, sep 2017.

\bibitem{Berman1975}
B~L Berman and S~C Fultz.
\newblock {Measurements of the giant dipole resonance with monoenergetic
  photons}.
\newblock {\em Reviews of Modern Physics}, 47(3):713--761, 1975.

\bibitem{Carlos2007}
C~A Bertulani.
\newblock {Excitation of soft dipole modes in electron scattering}.
\newblock {\em Phys. Rev. C}, 75(2):24606, feb 2007.

\bibitem{Berger1977}
R.~Berger.
\newblock Features of the giant e1 resonance.
\newblock In Carlo~Schaerf Sergio~Costa, editor, {\em Photo Nuclear Reaction,
  Lecture Notes in Physics}, chapter~3, pages 1--222. Springer-Verlag, 1977.

\bibitem{Gaffney2013}
L.~P. et~al. Gaffney.
\newblock Studies of pear-shaped nuclei using accelerated radioactive beams.
\newblock {\em Nature}, 497(7448):199--204, 2013.

\bibitem{Butler:2019qox}
P.~A. Butler et~al.
\newblock {The observation of vibrating pear-shapes in radon nuclei}.
\newblock {\em Nature Commun.}, 10(1):2473, 2019.

\bibitem{Butler:2020rmc}
P.~A. Butler et~al.
\newblock {Evolution of Octupole Deformation in Radium Nuclei from Coulomb
  Excitation of Radioactive $^{222}\mathrm{Ra}$ and $^{228}\mathrm{Ra}$ Beams}.
\newblock {\em Phys. Rev. Lett.}, 124(4):042503, 2020.

\bibitem{Chishti2020}
M.~M. R. et~al. Chishti.
\newblock Direct measurement of the intrinsic electric dipole moment in
  pear-shaped thorium-228.
\newblock {\em Nature Physics}, 16(8):853--856, 2020.

\bibitem{Behr:2022hym}
J.~A. Behr.
\newblock {Nuclei with enhanced Schiff moments in practical elements for atomic
  and molecular EDM measurements}.
\newblock 3 2022.

\bibitem{Bru2015}
Carl~R. Brune and Barry Davids.
\newblock Radiative capture reactions in astrophysics.
\newblock {\em Annual Review of Nuclear and Particle Science}, 65(1):87--112,
  2015.

\bibitem{Fri2019}
I.~Fri\ifmmode \check{s}\else \v{s}\fi{}\ifmmode \check{c}\else
  \v{c}\fi{}i\ifmmode~\acute{c}\else \'{c}\fi{}, T.~W. Donnelly, and R.~G.
  Milner.
\newblock New approach to determining radiative capture reaction rates at
  astrophysical energies.
\newblock {\em Phys. Rev. C}, 100:025804, Aug 2019.

\bibitem{Bet1934}
H.~{Bethe} and W.~{Heitler}.
\newblock {On the Stopping of Fast Particles and on the Creation of Positive
  Electrons}.
\newblock {\em Proceedings of the Royal Society of London Series A},
  146(856):83--112, August 1934.

\bibitem{Dav1952}
Charlotte~Meaker Davisson and Robley~D. Evans.
\newblock Gamma-ray absorption coefficients.
\newblock {\em Rev. Mod. Phys.}, 24:79--107, Apr 1952.

\bibitem{Roc1972}
G.~Roche, J.~Arnold, J.~Augerat, L.~Avan, M.~Avan, J.~Bonnet, J.~Fargeix,
  A.~Fleury, J.~Jousset, M.J. Parizet, and M.~Vialle.
\newblock First results of triplet production measurements near the threshold
  with a streamer chamber technique: Experimental set-up.
\newblock {\em Nuclear Instruments and Methods}, 103(3):533--544, 1972.

\bibitem{Das1939}
{Marques da Silva, Aurelio}.
\newblock Contribution \`a l'\'etude de la mat\'erialisation de l'\'energie.
\newblock {\em Ann. Phys.}, 11(11):504--547, 1939.

\bibitem{Shi1941}
Kenichi Shinohara and Mitio Hatoyama.
\newblock Pair production in the field of an electron.
\newblock {\em Phys. Rev.}, 59:461--461, Mar 1941.

\bibitem{Ogl1945}
W.~E. Ogle and P.~Gerald Kruger.
\newblock Pair electrons formed in the field of an electron.
\newblock {\em Phys. Rev.}, 67:282--285, May 1945.

\bibitem{Phi1949}
J.~A. Phillips and P.~Gerald Kruger.
\newblock On the production of electron pairs in the field of an electron.
\newblock {\em Phys. Rev.}, 76:1471--1478, Nov 1949.

\bibitem{Eps2016}
Charles~S. Epstein and Richard~G. Milner.
\newblock Qed radiative corrections to low-energy m\o{}ller and bhabha
  scattering.
\newblock {\em Phys. Rev. D}, 94:033004, Aug 2016.

\bibitem{Eps2020}
C.~S. Epstein et~al.
\newblock {Measurement of M\o{}ller scattering at 2.5 MeV}.
\newblock {\em Phys. Rev. D}, 102(1):012006, 2020.

\bibitem{Feng:2016jff}
Jonathan~L. Feng, Bartosz Fornal, Iftah Galon, Susan Gardner, Jordan Smolinsky,
  Tim M.~P. Tait, and Philip Tanedo.
\newblock {Protophobic Fifth-Force Interpretation of the Observed Anomaly in
  $^8$Be Nuclear Transitions}.
\newblock {\em Phys. Rev. Lett.}, 117(7):071803, 2016.

\bibitem{Holdom:1985ag}
Bob Holdom.
\newblock {Two U(1)'s and Epsilon Charge Shifts}.
\newblock {\em Phys. Lett. B}, 166:196--198, 1986.

\bibitem{BaBar:2014zli}
J.~P. Lees et~al.
\newblock {Search for a Dark Photon in $e^+e^-$ Collisions at BaBar}.
\newblock {\em Phys. Rev. Lett.}, 113(20):201801, 2014.

\bibitem{BaBar:2017tiz}
J.~P. Lees et~al.
\newblock {Search for Invisible Decays of a Dark Photon Produced in
  ${e}^{+}{e}^{-}$ Collisions at BaBar}.
\newblock {\em Phys. Rev. Lett.}, 119(13):131804, 2017.

\bibitem{Caputo:2021eaa}
Andrea Caputo, Alexander~J. Millar, Ciaran A.~J. O'Hare, and Edoardo
  Vitagliano.
\newblock {Dark photon limits: A handbook}.
\newblock {\em Phys. Rev. D}, 104(9):095029, 2021.

\bibitem{Antypas:2022asj}
D.~Antypas et~al.
\newblock {New Horizons: Scalar and Vector Ultralight Dark Matter}.
\newblock 3 2022.

\bibitem{McDermott:2017qcg}
Samuel~D. McDermott, Hiren~H. Patel, and Harikrishnan Ramani.
\newblock {Dark Photon Decay Beyond The Euler-Heisenberg Limit}.
\newblock {\em Phys. Rev. D}, 97(7):073005, 2018.

\bibitem{Pospelov:2008jk}
Maxim Pospelov, Adam Ritz, and Mikhail~B. Voloshin.
\newblock {Bosonic super-WIMPs as keV-scale dark matter}.
\newblock {\em Phys. Rev. D}, 78:115012, 2008.

\bibitem{Hanneke:2008tm}
D.~Hanneke, S.~Fogwell, and G.~Gabrielse.
\newblock {New Measurement of the Electron Magnetic Moment and the Fine
  Structure Constant}.
\newblock {\em Phys. Rev. Lett.}, 100:120801, 2008.

\bibitem{Pospelov_secluded_PhysRevD.80.095002}
Maxim Pospelov.
\newblock Secluded u(1) below the weak scale.
\newblock {\em Phys. Rev. D}, 80:095002, Nov 2009.

\bibitem{Pospelov_light_2017kep}
Maxim Pospelov and Yu-Dai Tsai.
\newblock {Light scalars and dark photons in Borexino and LSND experiments}.
\newblock {\em Phys. Lett. B}, 785:288--295, 2018.

\bibitem{Parker:2018vye}
Richard~H. Parker, Chenghui Yu, Weicheng Zhong, Brian Estey, and Holger
  M\"uller.
\newblock {Measurement of the fine-structure constant as a test of the Standard
  Model}.
\newblock {\em Science}, 360:191, 2018.

\bibitem{Morel:2020dww}
L\'eo Morel, Zhibin Yao, Pierre Clad\'e, and Sa\"\i{}da Guellati-Kh\'elifa.
\newblock {Determination of the fine-structure constant with an accuracy of 81
  parts per trillion}.
\newblock {\em Nature}, 588(7836):61--65, 2020.

\bibitem{Gardner:2019mcl}
Susan Gardner and Xinshuai Yan.
\newblock {Light scalars with lepton number to solve the $(g-2)_e$ anomaly}.
\newblock {\em Phys. Rev. D}, 102(7):075016, 2020.

\bibitem{Milstein:1994zz}
A.~I. Milstein and Martin Schumacher.
\newblock {Present status of Delbruck scattering}.
\newblock {\em Phys. Rept.}, 243:183--214, 1994.

\bibitem{Workman:2022ynf}
R.~L. Workman and Others.
\newblock {Review of Particle Physics}.
\newblock {\em PTEP}, 2022:083C01, 2022.

\bibitem{Berryman:2022zic}
Jeffrey~M. Berryman, Susan Gardner, and Mohammadreza Zakeri.
\newblock {Neutron Stars with Baryon Number Violation, Probing Dark Sectors}.
\newblock {\em Symmetry}, 14(3):518, 2022.

\bibitem{FileviezPerez:2022ypk}
Pavel Fileviez~Perez et~al.
\newblock {On Baryon and Lepton Number Violation}.
\newblock 7 2022.

\bibitem{Mohapatra:1980qe}
Rabindra~N. Mohapatra and R.~E. Marshak.
\newblock {Local B-L Symmetry of Electroweak Interactions, Majorana Neutrinos
  and Neutron Oscillations}.
\newblock {\em Phys. Rev. Lett.}, 44:1316--1319, 1980.
\newblock [Erratum: Phys.Rev.Lett. 44, 1643 (1980)].

\bibitem{Arnold:2012sd}
Jonathan~M. Arnold, Bartosz Fornal, and Mark~B. Wise.
\newblock {Simplified models with baryon number violation but no proton decay}.
\newblock {\em Phys. Rev. D}, 87:075004, 2013.

\bibitem{Gardner:2018azu}
Susan Gardner and Xinshuai Yan.
\newblock {Processes that break baryon number by two units and the Majorana
  nature of the neutrino}.
\newblock {\em Phys. Lett. B}, 790:421--426, 2019.

\bibitem{Weinberg:1979sa}
Steven Weinberg.
\newblock {Baryon and Lepton Nonconserving Processes}.
\newblock {\em Phys. Rev. Lett.}, 43:1566--1570, 1979.

\bibitem{Gardner:2017szu}
Susan Gardner and Xinshuai Yan.
\newblock {Phenomenology of neutron-antineutron conversion}.
\newblock {\em Phys. Rev. D}, 97(5):056008, 2018.

\bibitem{GYprep}
Susan Gardner and Xinshuai Yan.
\newblock {Prospects for the accelerator-based discovery of baryon-number
  violation by two units}.
\newblock to appear on the arXiv.

\bibitem{SVGproc}
Susan Gardner.
\newblock {New Opportunities for the Study of Baryon-Number Violation at
  Low-Energy Accelerators}.
\newblock contribution to these proceedings.

\bibitem{Phillips:2014fgb}
D.~G. Phillips, II et~al.
\newblock {Neutron-Antineutron Oscillations: Theoretical Status and
  Experimental Prospects}.
\newblock {\em Phys. Rept.}, 612:1--45, 2016.

\bibitem{Addazi:2020nlz}
A.~Addazi et~al.
\newblock {New high-sensitivity searches for neutrons converting into
  antineutrons and/or sterile neutrons at the HIBEAM/NNBAR experiment at the
  European Spallation Source}.
\newblock {\em J. Phys. G}, 48(7):070501, 2021.

\bibitem{BDX:2017jub}
M.~Battaglieri et~al.
\newblock {Dark matter search in a Beam-Dump eXperiment (BDX) at Jefferson Lab:
  an update on PR12-16-001}.
\newblock 12 2017.

\bibitem{Batell:2014mga}
Brian Batell, Rouven Essig, and Ze'ev Surujon.
\newblock {Strong Constraints on Sub-GeV Dark Sectors from SLAC Beam Dump
  E137}.
\newblock {\em Phys. Rev. Lett.}, 113(17):171802, 2014.

\bibitem{SHiP:2020noy}
C.~Ahdida et~al.
\newblock {Sensitivity of the SHiP experiment to light dark matter}.
\newblock {\em JHEP}, 04:199, 2021.

\bibitem{COHERENT:2019kwz}
D.~Akimov et~al.
\newblock {Sensitivity of the COHERENT Experiment to Accelerator-Produced Dark
  Matter}.
\newblock {\em Phys. Rev. D}, 102(5):052007, 2020.

\bibitem{MiniBooNEDM:2018cxm}
A.~A. Aguilar-Arevalo et~al.
\newblock {Dark Matter Search in Nucleon, Pion, and Electron Channels from a
  Proton Beam Dump with MiniBooNE}.
\newblock {\em Phys. Rev. D}, 98(11):112004, 2018.

\bibitem{Izaguirre:2013uxa}
Eder Izaguirre, Gordan Krnjaic, Philip Schuster, and Natalia Toro.
\newblock {New Electron Beam-Dump Experiments to Search for MeV to few-GeV Dark
  Matter}.
\newblock {\em Phys. Rev. D}, 88:114015, 2013.

\bibitem{Berlin:2018bsc}
Asher Berlin, Nikita Blinov, Gordan Krnjaic, Philip Schuster, and Natalia Toro.
\newblock {Dark Matter, Millicharges, Axion and Scalar Particles, Gauge Bosons,
  and Other New Physics with LDMX}.
\newblock {\em Phys. Rev. D}, 99(7):075001, 2019.

\bibitem{LDMX:2018cma}
Torsten \r{A}kesson et~al.
\newblock {Light Dark Matter eXperiment (LDMX)}.
\newblock 8 2018.

\bibitem{Snowmass-DP}
G.~Krnjaic and N.~Toro.
\newblock {Dark matter production at intensity frontier experiments}.
\newblock Snowmass RF6 summary white paper, to appear on arxiv.

\bibitem{Fabbrichesi_2021}
Marco Fabbrichesi, Emidio Gabrielli, and Gaia Lanfranchi.
\newblock {\em The Physics of the Dark Photon}.
\newblock Springer International Publishing, 2021.

\bibitem{Akesson:2022vza}
Torsten \r{A}kesson et~al.
\newblock {Current Status and Future Prospects for the Light Dark Matter
  eXperiment}.
\newblock In {\em {2022 Snowmass Summer Study}}, 3 2022.

\bibitem{Nelson:1989fx}
Ann~E. Nelson and Nikolaos Tetradis.
\newblock {Constraints on a New Vector Boson Coupled to Baryons}.
\newblock {\em Phys. Lett. B}, 221:80--84, 1989.

\bibitem{Tulin:2014tya}
Sean Tulin.
\newblock {New weakly-coupled forces hidden in low-energy QCD}.
\newblock {\em Phys. Rev. D}, 89(11):114008, 2014.

\bibitem{Berryman:2021jjt}
Jeffrey~M. Berryman and Susan Gardner.
\newblock {Neutron star structure with a new force between quarks}.
\newblock {\em Phys. Rev. C}, 104(4):045802, 2021.

\bibitem{Strumia:2021ybk}
Alessandro Strumia.
\newblock {Dark Matter interpretation of the neutron decay anomaly}.
\newblock {\em JHEP}, 02:067, 2022.

\bibitem{Fornal:2018eol}
Bartosz Fornal and Benjamin Grinstein.
\newblock {Dark Matter Interpretation of the Neutron Decay Anomaly}.
\newblock {\em Phys. Rev. Lett.}, 120(19):191801, 2018.
\newblock [Erratum: Phys.Rev.Lett. 124, 219901 (2020)].

\bibitem{KamLAND:2005pen}
T.~Araki et~al.
\newblock {Search for the invisible decay of neutrons with KamLAND}.
\newblock {\em Phys. Rev. Lett.}, 96:101802, 2006.

\bibitem{Benson}
Stephen Benson, Bogdan Wojtsekhowski, Barislav Vlahovic, and Serkan Golge.
\newblock {Opportunities and Challenges of a Low-energy Positron Source in the
  LERF}.
\newblock {\em AIP Conference Proceedings of International Workshop on Physics
  with Positrons at Jefferson Laboratory}, 1970:050004--1, 2018.

\end{thebibliography}
\smallskip

\end{document}